# Spin-Wave Dynamics in the Presence of Magnetic Vortices

Sławomir Mamica

**Abstract**

This chapter describes spin-wave excitations in nanosized dots and rings in the presence of the vortex state. The special attention is paid to the manifestation of the competition between exchange and dipolar interactions in the spin-wave spectrum as well as the correlation between the spectrum and the stability of the vortex. The calculation method uses the dynamic matrix for an all-discrete system, the numerical diagonalization of which yields the spectrum of frequencies and spin-wave profiles of normal modes of the dot. We study in-plane vortices of two types: a circular magnetization in circular dots and rings and the Landau state in square rings. We examine the influence of the dipolar-exchange competition and the geometry of the dot on the stability of the vortex and on the spectrum of spin waves. We show that the lowest-frequency mode profile proves to be indicative of the dipolar-to-exchange interaction ratio and the vortex stability is closely related to the spin-wave profile of the soft mode. The negative dispersion relation is also shown. Our results obtained for in-plane vortices are in qualitative agreement with results for core-vortices obtained from experiments, micromagnetic simulations, and analytical calculations.

## 1. Introduction

One of the hottest topics nowadays are small magnetic dots and rings with a thickness in a range of few tens of nanometers and the diameter ranging from one hundred nanometers to a few micrometers. A strong interest in such systems originates from their potential applicability as well as rich physics [1]. The physical properties of magnetic nanodots are related mostly to the concurrence of two types of magnetic interactions, namely exchange and dipolar ones. Usually, the coexistence of long- and short-distance interactions leads to new phenomena, such as surface and subsurface localization of the spin waves in layered magnetic systems [2, 3], opening of the band gaps in magnonic crystals [4, 5], or splitting the spin-wave spectrum into subbands in patterned multilayers [6, 7]. In the case of exchange and dipolar interactions, the situation is even more interesting due to competitive effects of these two schematically shown in **Figure 1**.

The favorable alignment of two magnetic moments (also called them spins) coupled via exchange interaction depends on the sign of the so-called exchange integral, $J$, regardless of their mutual position. If $J > 0$ the spins are parallel (ferromagnetic, FM, coupling) while for $J < 0$ the spins are antiparallel (antiferromagnetic, AFM, coupling). Dipolar coupling, on the other hand, depends on the mutual positions of spins being FM if the spins are aligned one after another and AFM for spins alongside one another (see **Figure 1**). As a result, the ferromagnetic exchange interaction forces parallel configuration of spins leading to the magnetic monodomain whereas pure dipolar interaction leads to the in-plane alignment of spins and so-called labyrinth magnetic structures [8]. Additionally, the dipolar interaction is a long range one and consequently very sensitive for size and shape of the sample while the exchange interaction is local. Thus, the competition between these two also depends on the size and shape of the system.

The concurrence of these to competitive interactions is the origin of the variety of possible magnetic configurations and leads to the occurrence of magnetic vortices in nanosized dots and rings [9–12]. In the vortex configuration, a magnetization component lying in the plane of the dot forms a closure state. Depending on the shape of the system, this in-plane magnetization can be realized as a circular magnetization in circular dots and rings or as a Landau state (closure domain configuration) in square rings, as shown in **Figure 2a**. In square dots, according to the simulations [13], the magnetic configuration is a mixture of these two states: along borders Landau state appears, which is the effect of the minimization of the surface magnetic charges, while in the central part of the dot, the magnetization is circular as a result of the tendency to decrease the (local) exchange energy. The area of circular magnetization is relatively small; therefore, in large-square dots, the Landau state prevails in the major part of the dot. However, in small dots, the circular in-plane magnetization fails to fit the geometry of the system only in minor corner regions.

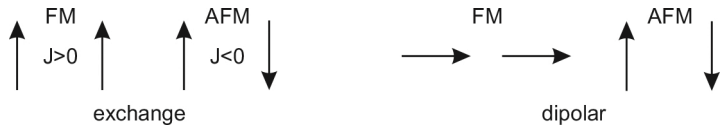

**Figure 1.** Exchange vs. dipolar interactions. Preferred configuration of magnetic moments depends on the sign of the exchange integral $J$ for the exchange interaction while on the alignment of magnetic moments for the dipolar interaction.

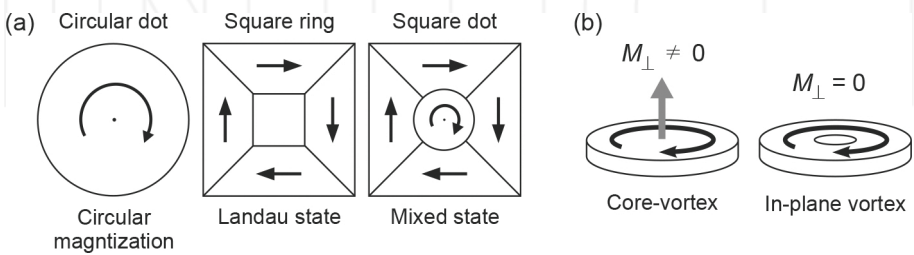

**Figure 2.** (a) Different preferred configurations of the in-plane magnetization component in dots of different shape. (b) Core-vortex vs. in-plane vortex.

For strong exchange interaction, the circular in-plane configuration is not enough to minimize the exchange energy at the vortex center (which is not necessary the dot center, however, for the stable vortex its center is in close vicinity to the center of the dot). As a consequence, spins at the center are rotated from their in-plane alignment (forced by dipolar interactions) forming so-called vortex core, a tiny region with a nonzero out-of-plane component of magnetization (**Figure 2b**). In typical ferromagnets, such as cobalt or permalloy, the exchange interaction is strong thus in experiments the vortex core is observed [14–16]. In rings, the center of the vortex is removed from the sample, thus the magnetization lies in the plane of the dot throughout its volume [17] except rings with extremely small internal radius [18]. The potential applications of the magnetic vortex itself increase from the possibility of the switching of core polarity (up or down) and chirality (the direction of the in-plane magnetization: clockwise, CW, or counterclockwise, CCW), and these two can be switched independently [19, 20].

In square dots, beside the vortex core, domain walls appear as well at the borders between domains. Roughly speaking, there are two types of domain walls: with and without nonzero out-of-plane magnetization (Bloch and Néel type, respectively) [21]. Thus, in the first case, the total out-of-plane magnetization is not zero even without the vortex core. Consequently, the out-of-plane magnetization can differ from zero in square rings in which the core does not appear. As we will show later, the preferred type of domain walls depends on the competition between exchange and dipolar interactions.

There are two types of magnetic excitations in magnetic dots in the vortex state. First one is a gyrotropic mode, i.e., the precession of the vortex core around the dot center. This is a low-frequency excitation with the frequency usually in the range of hundreds of MHz, and it can be utilized to microwave generation [22, 23]. The second type are spin waves; high-frequency excitations with the frequency of several GHz [24]. The spin-wave excitations are normal modes of the confined magnetic system similar to the vibration of the membrane. They prove to be of a key importance for the vortex switching [25], can be used to generate the higher harmonics of the microwave radiation [26], and have a significant influence on the vortex stability [27, 28].

In this chapter, we study the stability of the magnetic vortex state and the spin-wave excitations spectrum in two-dimensional (2D) nanosized dots and rings in their dependence on the competition between dipolar and exchange interactions. We use a very efficient method based on the discrete version of the Landau-Lifshitz equation. Our theoretical approach is described in Section 2. In next sections, we present our results starting with the circular dot in which the in-plane circular vortex is assumed as a magnetic state. In Section 3, we analyze an exemplar spin-wave spectrum of the dot showing typical effects such as the negative dispersion relation and the influence of the lattice symmetry on the spin-wave spectrum. In Section 4, we examine the stability of the in-plane vortex vs. the dipolar-to-exchange interaction ration (*d*) and the size of the dot. The influence of the competition between dipolar end exchange interactions on the spin-wave spectrum of a dot is studied in Section 5. In next two sections, we consider the influence of the spin-wave profile of the soft mode on the vortex stability in circular (Section 6) and square rings (Section 7). Finally, we provide some concluding remarks in Section 8.

## 2. The model

The object of our study is a dot (ring) cut out of a 2D lattice of elementary magnetic moments (**Figure 3**). For circular dots, the external size $L$ is defined as the number of lattice constants in the diameter of the circle used for cutting out the dot. The internal size of the ring, $L'$, is the radius of the inner circle (in units of the lattice constant). For square rings, $L$ means the number of lattice sites along the side of the square. Similarly, $L'$ means the side of the removed square. In linear approximation used in this work, the magnetic moment $M_R$, where $R$ is the position vector, can be expressed as a sum of two components: static, $M_{0,R}$, and dynamic, $m_R$, with the assumption that $|m_R| \ll |M_R|$, $|M_{0,R}| \simeq |M_R|$, and $m_R \perp M_{0,R}$. For any magnetic moment within the dot, we can define a local Cartesian coordinate system as follows: unit vector $i_R$ is parallel to the static component $M_{0,R}$, unit vector $j_R$ is oriented toward the vortex center lying in the plane of the dot, and unit vector $k_R$ is the third Cartesian unit vector being perpendicular to the other two. In this coordinate system, a dynamic component of the magnetic moment is $m_R = m_{j,R} j_R + m_{k,R} k_R$, where $m_{j,R}$ and $m_{k,R}$ we will refer to as in-plane and perpendicular coordinates of the magnetic moment, respectively. For in-plane vortices, the last component is always perpendicular to the plane of a dot.

The time evolution of any magnetic moment $M_R$ is described by the damping-free Landau-Lifshitz (LL) equation, which in the linear approximation reads:

$$i\frac{\omega}{\gamma\mu_0} m_R = M_{0,R} \times h_R + m_R \times H_R, \qquad (1)$$

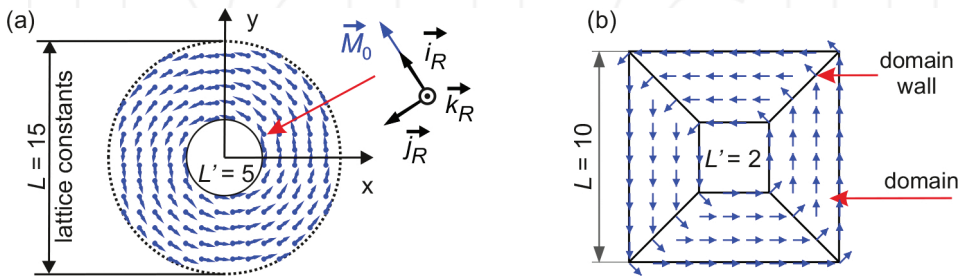

**Figure 3.** Schematic plots of two in-plane vortices typical for two types of rings: (a) a circular magnetization in a circular ring and (b) closure domains (Landau state) in a square ring. Both rings are based on a 2D square lattice with magnetic moments (represented by the arrows) arranged in the lattice sites. To the right in figure (a), the local coordinate system associated with the magnetic moment indicated by the arrow.

where $i$ is the imaginary unit, $\gamma$ is the gyromagnetic ratio, $\mu_0$ is the vacuum permeability, and $\omega$ is the frequency of harmonic oscillations of $m_R$. $H_R$ and $h_R$ are static and dynamic components of the effective field $H_R^{\text{eff}} = H_R + h_R$ acting on the magnetic moment $M_R$.

In this work, we consider exchange-dipolar systems only thus the effective field consists of two components:

$$H_R^{\text{eff}} = \frac{2J}{\mu_0(g\mu_B)^2} \sum_{R' \in \text{NN}} M_{R'} + \frac{1}{4\pi a^3} \sum_{R' \neq R} \left( \frac{3(R'-R)\left(M_{R'} \cdot (R'-R)\right)}{|R'-R|^5} - \frac{M_{R'}}{|R'-R|^3} \right).$$

The first term comes from the exchange interaction and can be derived from the Heisenberg Hamiltonian under the condition of uniform interactions. Since we restrict ourselves to nearest neighbor (NN) interactions the summation runs over NNs of the magnetic moment $M_R$. Here $J$ is the NN exchange integral, $\mu_B$ is the Bohr magneton, and $g$ is the g-factor. The second term is a typical dipolar sum over all magnetic moments within the sample except $M_R$. The position vectors $R$ are expressed in the units of the lattice constant $a$.

From Eq. (1) one can derive the system of equations of motion for dynamic components of all magnetic moments as follows:

$$i\Omega m_{r,R} = -\sum_{R' \in \text{NN}(R)} k_R \cdot m_{R'} + m_{k,R} \sum_{R' \in \text{NN}(R)} i_R \cdot i_{R'}$$

$$-d\left( \sum_{R' \neq R} \left( \frac{3[(R'-R) \cdot k_R][(R'-R) \cdot m_R]}{|R'-R|^5} - \frac{k_R \cdot m_{R'}}{|R'-R|^3} \right) + m_{k,R} \sum_{R' \neq R} \left( \frac{3[(R'-R) \cdot i_R][(R'-R) \cdot i_{R'}]}{|R'-R|^5} - \frac{i_R \cdot i_{R'}}{|R'-R|^3} \right) \right)$$

$$i\Omega m_{k,R} = \sum_{R' \in \text{NN}(R)} j_R \cdot m_{R'} - m_{r,R} \sum_{R \in \text{NN}(R)} i_R \cdot i_{R'}$$

$$+d\left( \sum_{R' \neq R} \left( \frac{3[(R'-R) \cdot j_R][(R'-R) \cdot m_R]}{|R'-R|^5} - \frac{j_R \cdot m_{R'}}{|R'-R|^3} \right) - m_{r,R} \sum_{R' \neq R} \left( \frac{3[(R'-R) \cdot i_R][(R'-R) \cdot i_{R'}]}{|R'-R|^5} - \frac{i_R \cdot i_{R'}}{|R'-R|^3} \right) \right), \quad (2)$$

where $\Omega = (g\mu_B\omega)/(2\gamma SJ)$ is the reduced frequency of a spin-wave excitation, $S$ is the spin (we assume that all spins within the dot are the same thus any magnetic moment is equal $M_R = g\mu_B S$), and $d$ is the only material parameter of the model referred to as a dipolar-to-exchange interaction ratio given by:

$$d = \frac{(g\mu_B)^2 \mu_0}{8\pi a^3 J}. \quad (3)$$

The above system of equations can be represented as an eigenvalue problem the matrix of which is called a *dynamic matrix*. The diagonalization of the dynamic matrix leads to the spectrum of frequencies and profiles of normal excitations of the dot. The spin-wave profile is a spatial distribution of the dynamic components of magnetic moments, i.e., the distribution of the amplitude of the magnetic moment precession. Dynamic components obtained from diagonalization are complex numbers with a phase shift $\pi/2$ between the real and imaginary part, which gives $T/4$ shift in time, where $T = 2\pi/\omega$ is a period of oscillations for a given mode.

Usually, the distribution of these components obtained for the same mode is similar and differ in the intensity only. Therefore, if the situation is clear, it is sufficient to provide one part (Re or Im) of the one component (in-plane or out-of-plane) to explain the character of the mode. The spin-wave profiles in circular dots are marked as (n, m) similarly to the vibrations of a membrane, i.e., accordingly to the number of nodal lines in the radial (n) and azimuthal (m) direction. The azimuthal modes occur in pairs: (n, −m) and (n, +m) with both modes of the same character; thus, in this work we denote them just (n, m), with m denoting |m|.

## 3. Spin-wave spectrum of a circular dot

In **Figure 4a** shows an example of the spin-wave spectrum obtained for a circular dot of the diameter $L = 101$. The dot is cut out from the square lattice and contains 8000 spins. A magnetic configuration is assumed to form an in-plane vortex. The spectrum is calculated for the dipolar-to-exchange interaction ratio $d = 0.42$. The shape of the spectrum is typical for exchange-dipolar systems: for low-frequency modes, the shape of the spectrum is determined by the dipolar interaction, while for the high frequencies by the exchange one. Of course, the spectrum is discrete, which is clearly seen in the inset where frequencies of 14 lowest modes are presented. Among these modes, one can distinguish a pairs of modes of the same frequencies. For example, modes 1 and 2, 7 and 8, 9 and 10, or 13 and 14 are degenerate in pairs. On the other hand, modes 3–6, 11 and 12 have unique frequencies.

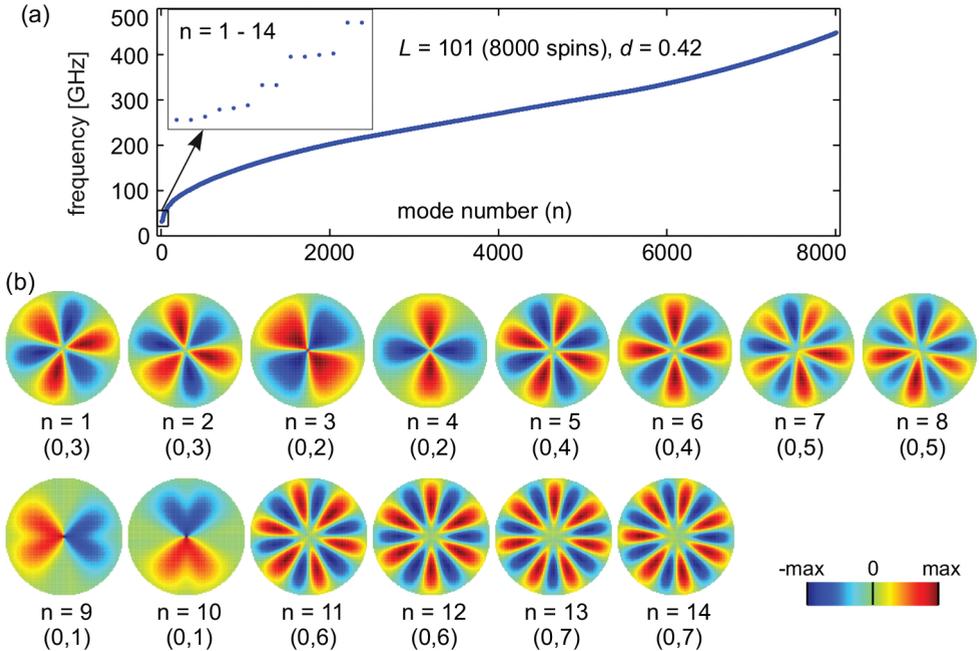

**Figure 4.** (a) An exemplar spin-wave spectrum calculated for the circular dot of the diameter $L = 101$ lattice constants consists of 8000 spins. The in-plane vortex configuration is assumed and the dipolar-to-exchange interaction ratio $d$ is set to 0.42. The inset shows 14 lowest modes of the spectrum. (b) Spin-wave profiles of 14 lowest modes of the spectrum shown in (a).

To investigate this feature, we provide spin-wave profiles of the lowest eight modes in **Figure 4b**. As we can see that these 14 modes include seven pairs of modes with the same absolute value of the azimuthal number. Degenerate modes are of the odd azimuthal number: (0,3) modes 1 and 2, (0,5) modes 7 and 8, (0,1) modes 9 and 10, and (0,7) modes 13 and 14. In contrast, for even azimuthal numbers, degeneration is lifted. This originates from the discreteness of the lattice the dot is cut out from. If the symmetry of the profile matches the symmetry of the lattice, the degeneration is removed. For example, mode 3 has two nodal lines coincide with high-spin density lines (along the $x$ and $y$ axes in **Figure 3a**). Its counterpart, i.e., mode 4 is rotated by $\pi/4$ having antinodal lines along the high-spin density lines. This situation is analogous to the boundary of the Brillouin zone in the periodic system where the energy gap appears between two excitations: one having nodes in the potential wells and the other one having antinodes. Indeed, if the dot is based on the square lattice it can be considered as a system periodic in the azimuthal direction. A unit cell in this case corresponds to a quarter of the dot and is delimited by high-spin density lines. In such picture, one-half of the wavelength of modes (0,2) fits the unit cell with nodes or antinodes at the unit cell boundary. The same rule holds for hexagonal lattice where the degeneration is lifted if the azimuthal number is divisible by 3 [29]. It is worth to noting that there is also another type of degeneracy lifting caused by the coupling of the azimuthal modes with the gyrotropic mode [30, 31] which is not related to the discreteness of the dot and appears even for first-order azimuthal modes. In our work, this is not the case since we assume coreless vortex as a magnetic configuration.

For the dot under consideration, radial and azimuthal numbers are related to the wave vector in the corresponding direction. Thus, the spectrum shown in **Figure 4** exhibits negative dispersion relation for modes (0,1), (0,2), and (0,3), i.e., for this modes, the frequency decreases with an increase of azimuthal number. Such negative dispersion was also observed for corevortices in circular dots experimentally [32, 33] and by means of analytical calculations [34, 35]. It was found that in a dot of a fixed thickness the increase in the diameter will cause the mode order to change, namely it will cause the negative dispersion to be stronger (the modes with higher azimuthal numbers will descend the spectrum). We show that this effect originates in the influence of the dipolar interaction regardless it is enhanced by the size of the system or by change of the dipolar-to-exchange interaction ratio.

## 4. Stability of the in-plane vortex

The dependence of the spin-wave spectrum on $d$ is shown in **Figure 5** for the dot under consideration ($L = 101$, 8000 spins). For intermediate values of the dipolar-to-exchange interaction ratio, there are no zero-frequency modes in the spectrum which means that the assumed in-plane vortex is a (meta)stable magnetic configuration (see, e.g., our discussion in reference [36]). Going toward smaller values of $d$ the exchange interaction gains the importance until $d = d_1$. From this point, the frequency of the lowest mode is zero and the in-plane vortex is no more stable (or even metastable); the lowest mode becomes the nucleation mode responsible for the reorientation of the magnetic configuration. The profile of this mode reflects the tendency of the system to find a new stable state. Since this transition is forced by the exchange interaction we will call it the exchange-driven reorientation (transition). While $d$ increases, which means the dipolar interaction gains the importance, another transition appears for $d = d_2$. In this case, the reorientation is caused by prevailing dipolar interaction; thus, it is referred to as dipolar-driven reorientation (transition). This behavior reflects the origin of the vortex state: the competition between dipolar and exchange interaction.

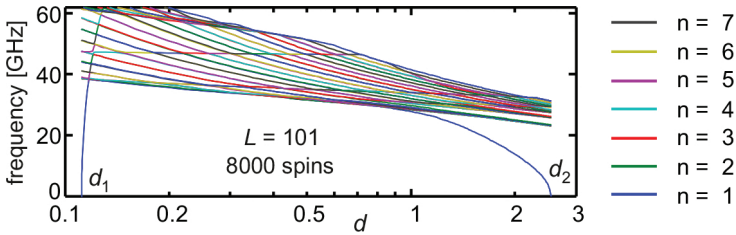

**Figure 5.** The frequency dependence on the dipolar-to-exchange interaction ratio $d$ (in logarithmic scale) for 36 lowest modes in the spin-wave spectrum of the circular dot of the diameter $L = 101$ in the in-plane vortex state. The color assignment is indicated at the right; the colors repeat cyclically for successive modes. There are no zero-frequency modes between two critical values $d_1$ and $d_2$ which is indicative for the stability of the assumed magnetic configuration.

The importance of the dipolar interaction depends, besides its dependence on $d$, also on the size of the system. Therefore, the critical values of $d$ should change with the dot size. **Figure 6a** shows critical values $d_1$ and $d_2$ vs. the number of spins, $N$, in which the dot consists of (which is equivalent to change of the dot diameter since the system is 2D). The critical value $d_2$ (for the dipolar driven reorientation) clearly depends on $N$, especially for small dots. Surprisingly, for the exchange-driven reorientation the critical value $d_1 = 0.1115$ and is constant in the whole range of the dot size, i.e., from 60 to 8000 spins ($L = 9–101$). (The same value and behavior of $d_1$ is reported in reference [37] where circular dots are studied by means of Monte Carlo simulations.)

To address this behavior of critical values in **Figure 6b**, we provide profiles of the lowest mode for two values of $d$, for $d \approx d_1$ (left profile) and for $d \approx d_2$ (right profile), for $L = 23$ (408 spins). Both profiles are localized at the vortex center but the localization near $d_1$ is much stronger than for $d_2$. The reorientation at $d_1$ is forced by the exchange interaction which is local and thus the dynamic interaction (between dynamic components of magnetic moments), confined to the very center of the dot, is not sensitive to the size of the dot. (It is not sensitive to the shape of the dot as well [28]). The second transition ($d_2$) is forced by the long-range dipolar interaction and the dynamic interaction of this type, although localized near the center, "feels" the dot size even for rather big dots. However, due to the localization this effect fades for larger dots, which is reflected in the $d_2$ curve in **Figure 6a**.

For typical ferromagnetic materials, the dipolar-to-exchange interaction ratio has very small value due to strong exchange. For example, using experimental data for ultrathin cobalt film

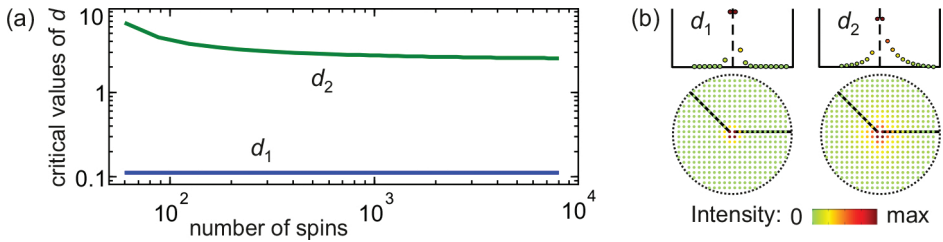

**Figure 6.** (a) Critical values $d_1$ and $d_2$ vs. the dot size (the number of magnetic moments within the dot, in logarithmic scale) for circular dot in the in-plane vortex state. (b) Exemplar profiles of the lowest mode in circular dots for $d≈d_1$ (left profile) and for $d≈d_2$ (right profile). Above each profile, its section along the indicated lines.

[38] from the relationship (3) we obtain $d_{Co} = 0.00043$ which is far below $d_1$. Consequently, in such materials, the in-plane vortex is unstable regardless the size of the dot (since $d_1$ is size independent).

## 5. Competition between interactions

As seen from **Figure 5**, for the majority of modes the frequency decreases with increasing $d$ but with different rate. As a result, the order of modes in the spectrum changes with $d$; this effect is particularly intensive at the bottom of the spectrum. In particular, the mode of the lowest frequency has different symmetry of its profile in different ranges of $d$ (compare **Figure 7**). The modes with the decreasing frequency can be divided into two groups: first one contains purely azimuthal modes (radial number equal zero). Within this group, the rate of the decreasing frequency grows with the increasing azimuthal number. However, above c.a. 55 GHz, this rate is visibly lower for another group of modes with the radial number 1. Within this second group, the situation repeats: for the mode (1,m) the frequency decrease rate is almost the same as for the mode (0,m) and it grows with increasing m. It shows that the impact of the dipolar-to-exchange interaction ratio on the mode frequency is determined mainly by its azimuthal number, the radial number being of little influence.

Besides the localized mode, there is one more mode in **Figure 5** the frequency of which acts in different way than the majority; in the broad range of $d$ its frequency is almost constant. This mode, called fundamental mode, is an analogue of the uniform excitation [35]. Its profile is almost uniform within the dots without any nodal lines in azimuthal nor a radial direction thus it is labeled as (0,0). Highly uniform profile is the origin of the independence of the frequency of the fundamental mode on $d$.

As we already noticed that the mode order in the spin-wave spectrum is influenced by the dipolar-to-exchange interaction ratio and by the size of the dot, thus influences the character of the lowest mode. **Figure 7a** shows the dependence of the lowest mode frequency on $d$ for different size of the dot. **Figure 7b** and c provides mode profiles for some values of $d$ for two dot diameters: 51 and 101, respectively. Close to the critical value $d_1$ the profile is strongly localized at the vortex center regardless the dot size. This strong localization together with the short range of the exchange interaction (responsible for the magnetic reorientation below $d_1$) results in not only the independence of $d_1$ on the size and shape of the dot but also the frequency vs. $d$ dependence is the same for the dot of any size. In this range of the dipolar-to-exchange interaction ratio, the lowest mode is a soft mode but with growing $d$ its frequency increases rapidly and the mode ascend the spectrum very fast causing crossings with azimuthal modes of decreasing frequencies. After the first crossing, the azimuthal mode becomes the lowest in the spectrum.

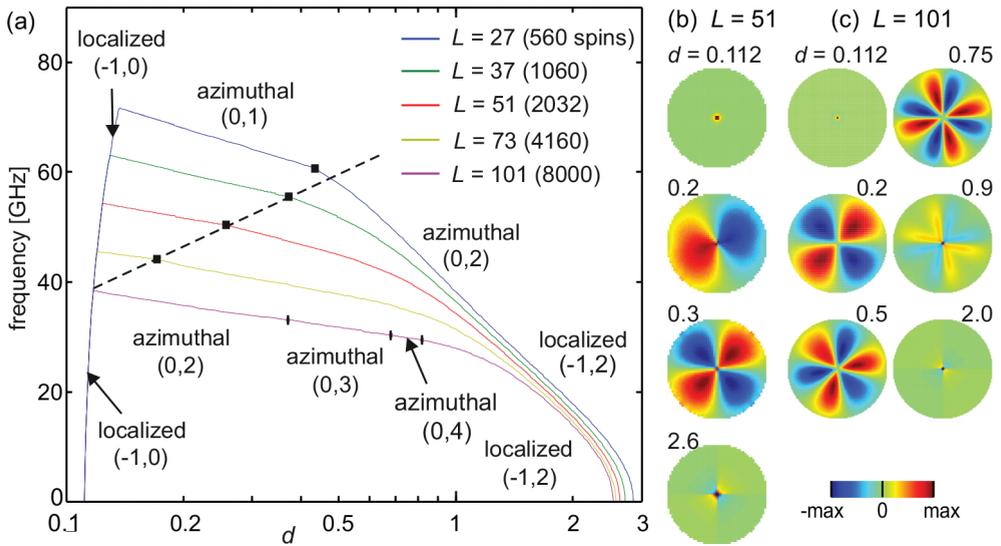

**Figure 7.** (a) The dependence of the lowest mode frequency vs. dipolar-to-exchange interaction ratio $d$ (in logarithmic scale) in circular dots of different diameter $L$ with the in-plane vortex as a magnetic configuration. On every curve, the crossing between first- and second-order azimuthal modes is marked with a black square (if exists). (b, c) Evolution of the lowest mode profile with $d$ in dots of diameters 51 and 101, respectively.

For small dots ($L < 100$), the mode (0,1) is the lowest one after crossing with the localized mode. While $d$ continues to increase till next crossing appears and (0,2) mode becomes the lowest. The point of crossing of these modes shifts to the smaller $d$ with increasing size of the dot (see **Figure 7a**). Finally, for $L = 101$, the crossing between modes (0,1) and (0,2) takes place for lower $d$ than the crossing with the localized mode. In a consequence, the first-order azimuthal mode is not the lowest one for any $d$. On the other hand, higher order modes may have the lowest frequency while $d$ is growing (compare **Figure 7b** and **c**).

Here, we observe a general tendency of two interactions in question. The dipolar interaction favors higher order azimuthal modes. Thus, modes with the increasing azimuthal number m fall successively to the bottom of the spectrum as this type of interaction gains in importance regardless of whether their strenghten is due to the size ($L$) or material ($d$) of the dot. The exchange interaction in contrast favors modes with m = 1. Thus, the competition between the exchange and dipolar interaction manifests itself not only in the preferred magnetic configuration but also in the profile of the lowest frequency modes.

This rule changes if the vortex is close to unstable, i.e., close to the critical value of $d$. In this case, the soft mode is strongly localized at the vortex center. But even here this localized mode has nodal lines in the azimuthal direction for strong dipolar interaction and uniform for strong exchange interaction.

## 6. Circular rings

In circular rings, the central part of the dot is removed along with the vortex center. This causes significant reduction in the influence of the exchange interaction and consequently should result in the stabilization of the in-plane vortex for lower values of the dipolar-to-exchange interaction ratio. **Figure 8a** shows the typical dependence of the spin-wave spectrum in circular rings on $d$. The exemplar ring has external diameter $L = 25$ and internal one $L' = 2$ which means that only four central magnetic moments are removed from the dot. The overall character of the picture is very similar to that for the dot shown in **Figure 5** with two exceptions: the range of the in-plane vortex stability and the behavior of the soft mode above $d_1$. (The decreasing of the frequency with growing $d$ is much faster mostly due to the smaller external diameter.)

Just above $d_1$ the frequency of the soft mode increases steeply, as a consequence of increasing stability of the system, but before first crossing with the azimuthal mode the frequency slows down and finally becomes almost independent on $d$. The profile of this mode is shown in **Figure 8b** for $d = 0.01$; it is no more localized. Instead of this, the mode is a fundamental mode (0,0) being almost uniform within the ring. Due to the lack of the topological defect, there is no reason for the localization.

Other profiles provided in **Figure 8b** illustrate the change of the character of the lowest mode. Even if the external diameter of the ring is rather small, higher order azimuthal modes are the lowest for large enough $d$: (0,3) for $d = 1.3$ and (0,4) for $d = 2.0$. In full dots these modes could be the lowest for c.a. 4 times larger diameter which reflects the change of the balance between exchange and dipolar interaction after removing only few magnetic moments from the center of the dot.

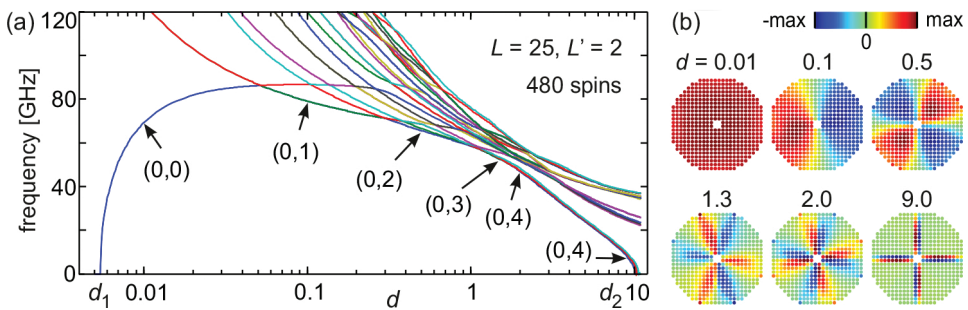

**Figure 8.** (a) The frequency dependence on the dipolar-to-exchange interaction ratio $d$ (in logarithmic scale) for 25 lowest modes in the spin-wave spectrum of the circular ring of the external diameter $L = 25$ and the internal one $L' = 2$ in the in-plane vortex state. (b) The evolution of the lowest mode profile. Profiles are calculated for six values of $d$ marked with arrows in (a).

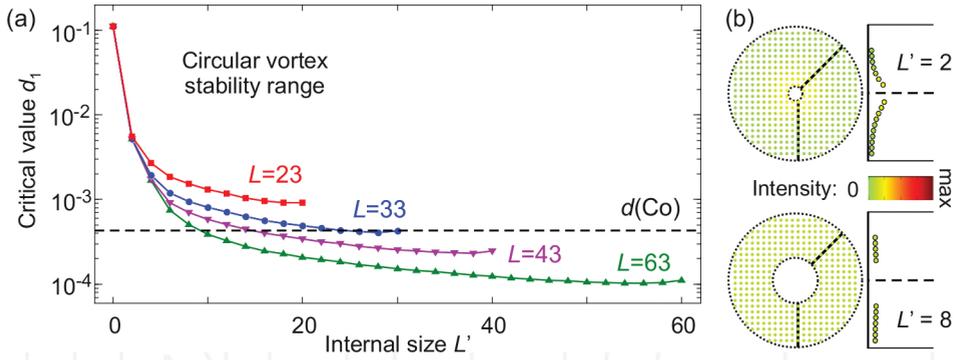

**Figure 9.** (a) Critical value $d_1$ vs. the internal diameter $L'$ of the circular ring for four external diameters $L$. The dashed line in (a) indicates the value of $d$ for Co/Cu(001) calculated from experimental results reported in reference [38]. (b) Spin-wave profiles of the soft mode in circular rings calculated for $d≈d_1$ for two internal diameters $L'$. The external diameter of rings is fixed at $L = 23$. To the right of each profile, its section along the indicated lines.

The removing of these four central magnetic moments has also great impact on the stability of the in-plane vortex, as it should be expected. The critical value $d_1$ decreases from 0.115 for the full dot down to 0.0052 for the ring under consideration. However, this new critical value is still much larger than the value of $d$ in common ferromagnetic materials. **Figure 9a** shows the change of the critical value $d_1$ with increasing the internal diameter of the ring for four external diameters: 23, 33, 43, and 63. In contrast to full dots in the rings $d_1$ visibly depends on both internal and external diameters (though for very small internal diameter the influence of the external size is weak). The increase in any diameter of the ring enhances the stability of the in-plane vortex. As a result, this magnetic configuration is stable even for such a material as cobalt if the ring is large enough ($d_1 < d_{Co}$).

The enhancement of the in-plane vortex stability due to the increasing of its internal diameter is rather obvious if we notice that the local exchange interaction between neighboring magnetic moments increases with decreasing distance from the vortex center (due to the change in the angle between them). In this context, the removal of the bigger circle from the center of the dot means the decreasing of the exchange interaction at the internal edge of the ring. Of course, this change in the exchange energy at the border should be visible in spin-wave profiles. To illustrate this effect, we calculate the profiles of the lowest mode for $d≈d_1$ for the ring of the external diameter $L = 23$ and two different internal diameters, $L' = 2$ and $L' = 8$, shown in **Figure 9b**. Successive removing of the central part of the dot results in decreasing of the amplitude of the magnetic moments precession (smaller intensity of the profile) at the internal edge of the ring. On the other hand, the amplitude is slightly increased in the rest of the ring, especially at the outer edge. For larger hole in the ring, the profile is almost uniform in radial direction and $d_1$ is very little dependent on $L'$. This nonzero intensity of the spin-wave profile

## 7. Square rings

In square rings, the in-plane vortex takes the form of the Landau state (closure domain configuration, see **Figure 2**). Unlike circular rings, here the magnetization along internal and external edges has the same conditions (no curvature). Another difference is the existence of domains walls. To see how these dissimilarities influence the in-plane vortex stability in **Figure 10a** we show the critical value $d_1$ vs. the internal size $L'$ for square rings of different external size $L$. Similarly to the circular rings, the removal of the central part of the dot results in the drop of $d_1$, i.e., the in-plane vortex becomes stable for stronger exchange interaction. The critical value changes from 0.115 to 0.049. This time, in contrast to the previous case, this value is constant for broad range of the internal size of the ring. Additionally, $d_1$ does not depend on the external size of the ring as well. Therefore, the in-plane vortex (with the domain walls of Néel type) is not stable in square rings made from typical ferromagnetic materials.

To explain this behavior, **Figure 10b** shows spin-wave profiles of the lowest mode for $d \approx d_1$ for square rings of the external size $L = 22$ and three different internal sizes: $L' = 0$ (full dot), $L' = 2$, and $L' = 16$. Removing of the central part of a dot, even just few magnetic moments, destroys the central localization of the lowest mode as it was in the case of circular dots, but now the localization is shifted to the corners of the resultant ring. Such corner-localized profile is not affected by the change of the size of the ring in large range of both, internal and external size. Again, the strongly localized spin-wave profile together with the local character of the exchange interaction causes the critical value of $d$ for the exchange-driven reorientation to be independent on the size of the system. The high amplitude of the spin wave at the corners suggests also the increasing of the out-of-plane component of magnetization which means the formation of the Bloch-type domain walls.

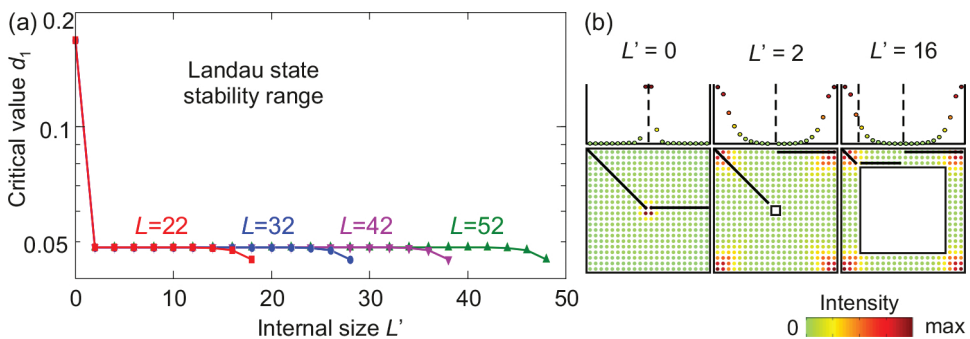

**Figure 10.** (a) Critical value $d_1$ vs. the internal size $L'$ of the square ring for four external sizes $L$. (b) Spin-wave profiles of the soft mode in square rings calculated for $d \approx d_1$ for three internal sizes $L'$. The external size of rings is fixed at $L = 22$. Above each profile, its section along the indicated lines.

## 8. Concluding remarks

In this chapter, we have shown our results concerning spin wave normal modes in nanosized dots and rings in the presence of the in-plane magnetic vortex. In experiments, in-plane vortices are observed in rings while in full dots made from typical ferromagnetic materials (e.g., cobalt or permalloy) the vortex core is formed at the vortex center [30]. Our results obtained for circular dots are consistent with this observation: in-plane vortex is stable in such a system for very weak exchange interaction, much weaker than in usual ferromagnets. We obtain the critical dipolar-to-exchange interaction ratio $d_1 = 0.1115$ (which corresponds to the exchange integral $J = 0.058$ eV) and this value is the same as received from Monte Carlo simulations in reference [37]. This critical value does not depend on the size of a dot which is also in agreement with simulations [37].

An interesting finding is the stability of the in-plane vortex in rings. In circular rings, the removal of a central part of a dot brings the dependence of $d_1$ on both diameters of the ring (external and internal) and, consequently, the in-plane vortex becomes stable even for strong exchange if the ring is large enough. In square rings, the situation is completely different: $d_1$ does not depend on any size of the ring (except extremely narrow rings). The critical value $d_1$ is reduced in comparison with full dots though not enough to stabilize the in-plane vortex. Therefore, in square rings made from usual ferromagnetic materials, the in-plane vortex is not stable (due to the preferred type of domain walls).

For the in-plane vortex configuration in full dots we found similar phenomena as reported from experiments, micromagnetic simulations, and analytical calculations, except those which arise from the existence of the gyrotropic motion of the vortex core, e.g., the splitting of the spin-wave frequency due to the coupling to the gyrotropic mode [31]. The qualitative agreement between results for in-plane and core vortices is an effect of the existence of the vortex center. Even without the out-of-plane component of the magnetization, the center of the vortex plays the role of the topological defect in the same manner as the vortex core. This defect acts as a nucleation center if $d$ reaches its critical value and cause the localization of the soft mode. On the other hand, the properties of the spin waves in the presence of the vortex originate from the competition between exchange and dipolar interaction; thus the effects such as negative dispersion relation or diversity of the lowest mode profiles are similar for both types of vortices: with and without the core.

In our model, the dot is cut out from a discrete lattice which obviously has a consequence in the results. If the symmetry of azimuthal modes matches the symmetry of the lattice, the frequency of modes with opposite azimuthal numbers splits. Also the fundamental mode, an analogue of the uniform excitation, has nonuniform spin-wave profile whose symmetry reflects the symmetry of the lattice. (A similar effect was observed in micromagnetic simulations due to the artificial discretization of a sample [39–41].) In the case of circular dots and rings based on the discrete lattice, the edges are not smooth circles and cannot be smoothed as it is in continuous systems with artificial discretization, e.g., in micromagnetic simulations [42]. With the size of the ring, the edge smoothness increases but even for rather small dots (a dozen of lattice constants in the diameter) we obtain self-consistent results.

In this work, the method described in Section 2 is used for 2D dots and rings but its applicability is far beyond these simple systems. It can be used for 2D or 3D systems of an arbitrary shape, size, lattice, or magnetic configuration. Moreover, if the exchange interaction is neglected the method can be applied for nonperiodic systems too. Also interactions taken into account are not limited to dipolar and exchange only (the model with the anisotropy and the external field taken into account is derived in reference [43]). The main disadvantage of our approach is the lack of simulations; the assumption, instead of the simulation, of the magnetic configuration is useful for very simple magnetic configurations only. On the other hand, in comparison with time-domain simulations, the time of calculations is very short, and the spin-wave spectrum is obtained directly from diagonalization of the dynamic matrix (without the usage of the Fourier transformation). For simple magnetic configurations, our results are in perfect agreement with simulations [37, 13, 44]. In the case of more complicated systems, the simulations should be used for finding the stable magnetic configuration and for the simulated configuration the dynamical matrix method can be used to obtain the spin-wave spectrum.

# Acknowledgements

The author thanks Jean-Claude Serge Lévy and Maciej Krawczyk for valuable discussions. The author received fund from Polish National Science Centre project DEC-2-12/07/E/ST3/00538 and from the EUs Horizon2020 research and innovation program under the Marie Sklodowska-Curie GA No644348.

# Author details

Sławomir Mamica

Address all correspondence to: mamica@amu.edu.pl

Faculty of Physics, Adam Mickiewicz University in Poznan, Poznan, Poland